\documentclass[amsmath,amssymb,aps,floatfix,prl,superscriptaddress,twocolumn]{revtex4-1}

\usepackage{graphicx}
\usepackage{hyperref}
\begin{document}	
	\title{Diffusion in liquid metals is directed by competing collective modes }
	
	\author{Franz Demmel}
	\email{franz.demmel@stfc.ac.uk}
	\affiliation{ISIS Facility, Rutherford Appleton Laboratory, Didcot, OX11 0QX, UK}	
	\author{Noel Jakse}
	\affiliation{Univ. Grenoble Alpes, CNRS, Grenoble INP, SIMaP, F-38000 Grenoble, France}
	
\begin{abstract}
The self-diffusion process in a dense liquid is influenced by collective particle movements.
Extensive molecular dynamics simulations for liquid aluminium and rubidium evidence a crossover in the diffusion coefficient at about $1.4$ times the melting temperature $T_m$, indicating a profound change in the diffusion mechanism.
The corresponding velocity auto-correlation functions demonstrate a decrease of the cage effect with a gradual set-in of a power-law decay, the celebrate {\it long time tail}. This behavior is caused by a competition of density fluctuations near the melting point with vortex-type particle patterns from transverse currents in the hot fluid. The investigation of the velocity autocorrelation function
evidences a gradual transition in dynamics with rising temperature.
The competition between these two collective particle movements, one hindering and one enhancing the diffusion process,
leads to a non-Arrhenius-type behavior of the diffusion coefficient around $1.4~T_m$, which signals the transition from a dense to a fluid-like liquid dynamics in the potential energy landscape picture.
	
\end{abstract}
	
\date{\today}
	
\maketitle
Mass transport is described by the self-diffusion coefficient and plays an important role in phenomena ranging from active matter, \textit{e.g.} from ion transport in living cells \cite{Li2024,nolte2024} to crystallization in material science \cite{sosso2016}.
A profound understanding of particle transport in the liquid state remains a challenging task.
In particular, metals show a wide temperature range of the liquid state, in which the character of mass transport might display a peculiar behavior.
	
Within the hydrodynamic limit, which means long times compared to collision times of the particles and
long distances compared to typical length scales between collisions, the change of the tagged particle density with space and time can be described by a partial differential equation \cite{hansen2006}.
The solution is a Gaussian distribution for the particle density, which is used in tracer diffusion experiments to determine the diffusion coefficient. Further experimental methods to determine diffusion coefficients are, for example, nuclear magnetic resonance, dynamic light scattering and quasielastic neutron scattering, but these measurements remain challenging.
Within molecular dynamics (MD) simulations, the diffusion coefficient can be obtained through the long time limit of
the mean-square displacement of the particles as well as through the time integral of the velocity autocorrelation function (VACF), a Green-Kubo relation between a correlation function and a transport parameter \cite{hansen2006}.
	
In low-density fluids kinetic theory can predict the diffusion coefficient quite precisely
starting with the Boltzmann equation and then incorporating particle correlations within the Enskog theory.
That description leads to an exponentially decaying VACF, $\psi(t)\propto exp(-\Gamma_E t)$, with the Enskog collision parameter $\Gamma_E $\cite{boon1980,balucani1994}.
	
With increasing density correlated collisions become more and more important and need to be taken into account.
Early MD-simulations on moderate dense hard-sphere systems by Alder and Wainwright revealed an algebraic decay of the VACF at long times, the so-called {\it long time tail} (LTT), namely $\psi(t) \propto t^{-3/2}$  \cite{alder1967,alder1970,dorfman1970}. This power-law decay of the VACF contributes to an increased diffusion coefficient and was then confirmed on Lennard-Jones generic models \cite{levesque1974}.
In a simple picture, the enhancement of the diffusion coefficient is caused by vortex-type particle movements, which could be understood from hydrodynamics \cite{alder1970,zwanzig1970}. The shear modes generated by a moving particle act back on the particle at a later time. Within kinetic theory, ring-type collision events allow to correlate momentum over several mean free paths in the fluid and are at the origin for the LTT \cite{dorfman1970,ernst1971}. Mode coupling theory (MCT) is well adapted to describe these slow relaxation events and has been applied to describe the single particle motion \cite{dorfman1972,schepper1979,goetze1978,bosse1979,montfrooij1986}.
	
Experimentally, the LTT should appear in the density of states as a low frequency square root
dependence \cite{hansen2006}. For liquid sodium, it was shown that at temperatures
larger than twice the melting point the neutron scattering data are compatible with this prediction \cite{morkel1987}.
It was noted that with increasing density, the long time tail contribution will be eventually
suppressed by density fluctuations to a few percent of the signal \cite{bosse1979}.
Nevertheless, its rather weak existence could be revealed in a MD simulation towards the supercooled state \cite{williams2006}.
	
Increasing the density further, a different process of correlated particle movements appears, the so-called {\it cage effect}, for which the VACF shows a negative contribution after a first fast decay.
In a dense liquid, the tagged particle cannot immediately escape the cage of the surrounding neighbors and might be backscattered before moving out of the cage. This negative contribution in the VACF leads to a reduction in the diffusion coefficient below the expected Enskog value \cite{hansen2006}.
Within MCT, this backscattering effect can be described quantitatively through the inclusion of density fluctuations, which are ultimately responsible for the decay of the surrounding cage and enabling the diffusive motion \cite{sjogren1980,wahnstrom1982,balucani1994}. A quantitatively excellent agreement was achieved for the diffusion process of simple liquid metals near the melting point \cite{balucani1990,balucani1992}.
Diffusion coefficients of a hard-sphere fluid were simulated over a wide density range and showed a crossover from a diffusion coefficient smaller than $D_E$ to a larger one \cite{erpenbeck1991,hansen2006}.
A study on liquid sodium covering a wide range in densities over coarse steps confirms this picture from hard-sphere simulations \cite{balucani1995,pilgrim2006}. The crossover from a diffusion coefficient smaller than $D_E$ to a larger one occurs at a small density change, which amounts to only $3$\% for liquid Rb \cite{ohse1985}.
	
In this Letter, the temperature evolution of self-diffusion in liquid Al and Rb is investigated by means of MD-simulations for states spanning from the undercooled liquid to the supercritical fluid, representing a wide temperature and density range. Through machine learning a potential was derived which reaches the accuracy of first principles approaches and then enables a comprehensive study of the single particle dynamics over a wide temperature in unprecedented detail.
The diffusion coefficients show a crossover around $1.4~T_m$ which is caused by changes in the underlying collective particle dynamics.

Liquid Rb and Al were chosen as typical cases of monovalent and polyvalent metals, respectively, and with significantly different masses and melting points. Atomic interactions for Al were described through a machine learning potential that has proven to have close to \textit{ab initio} accuracy, especially in the liquid state for the dynamics \cite{jakse2023}(please see fig. 1 in the Supplemental Material for a comparison of diffusion coefficients \cite{SM}). For Rb an Embedded Atom Model (EAM) in the Finnis-Synclair form able to well reproduce the liquid state was used \cite{nichol2016}.
Isobaric-isothermal ensemble simulations (constant number of atoms $N$, pressure $P$, and temperature $T$) were carried out with the \textsc{Lammps} code \cite{Lammps} with $N=10976$ and $16000$ respectively for Al and Rb, these $N$ values being sufficiently large to prevent finite size effects \cite{Yeh2004}. The simulated diffusion coefficients were calculated from $\psi(t)$ using a Green-Kubo relation \cite{hansen2006}:
\begin{equation}
	D=\int_0^\infty \psi(t)dt=\frac{1}{3} \int_0^\infty dt~ {\langle \mathbf{v}(t+t_0)\cdot\mathbf{v}(t_0) \rangle}_{t_0}.
\end{equation}
Enskog diffusion coefficients $D_E$ for liquid Rb were obtained from experimental data.
It was suggested that density fluctuation relaxations at the structure factor maximum $Q_0$, where the deGennes narrowing occurs \cite{degennes1959}, can be interpreted as a first step in a diffusion process \cite{cohen1987}.
A relation connects the linewidth $\Gamma(Q_0)$ and $D_E$: $\Gamma=D_E Q^2 d(Q \sigma)/S(Q)$, where $d(Q \sigma)= (1-j_0(Q \sigma)+2j_2(Q \sigma))^{-1}$ is given by a combination of spherical Bessel functions $j_l$ and $\sigma$ denoting a hard sphere diameter. This relation provides access to $D_E$ through a scattering experiment on the collective dynamics. In contrast, for Al $D_E$ values were obtained through calculation according to \cite{balucani1994}: $D_E=(3/8~\sqrt{\pi})\sqrt{k_B T/m}/n \sigma^2 g(\sigma)$ with $\sigma=2.5~\AA$ the hard sphere diameter \cite{demmel2011} of an Al atom and $g(\sigma)$ the pair distribution function at contact.
It is well approximated by: $g(\sigma)=(1-\eta/2)/(1-\eta)^3$ with the packing fraction $\eta= \pi n \sigma^3/6$, the particle density $n$ being derived from the density \cite{assael2006}. These values are consistent with those studied previously \cite{jakse2013b}.
	
\begin{figure}
\includegraphics[angle=0,width=0.48\textwidth]{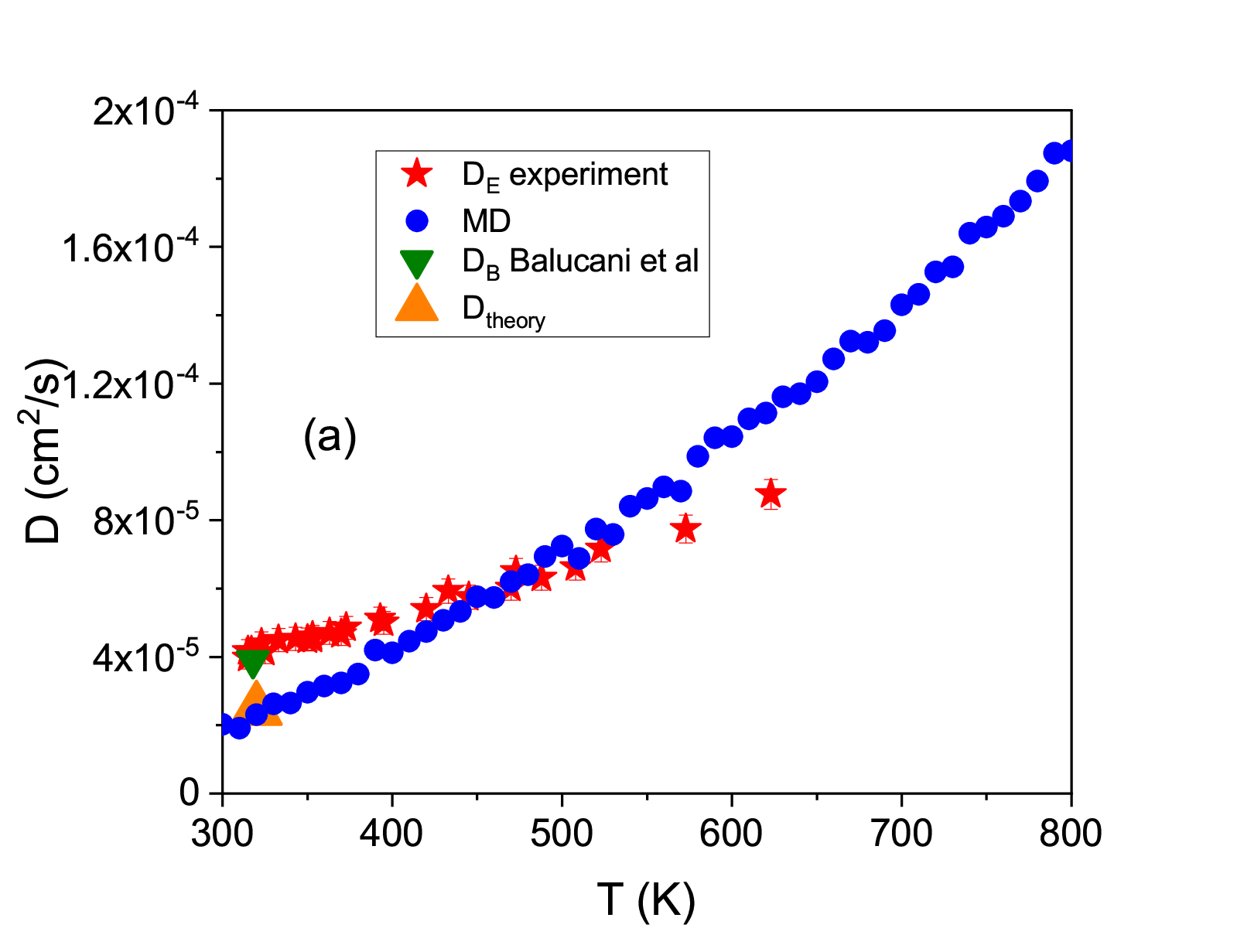}
\includegraphics[angle=0,width=0.48\textwidth]{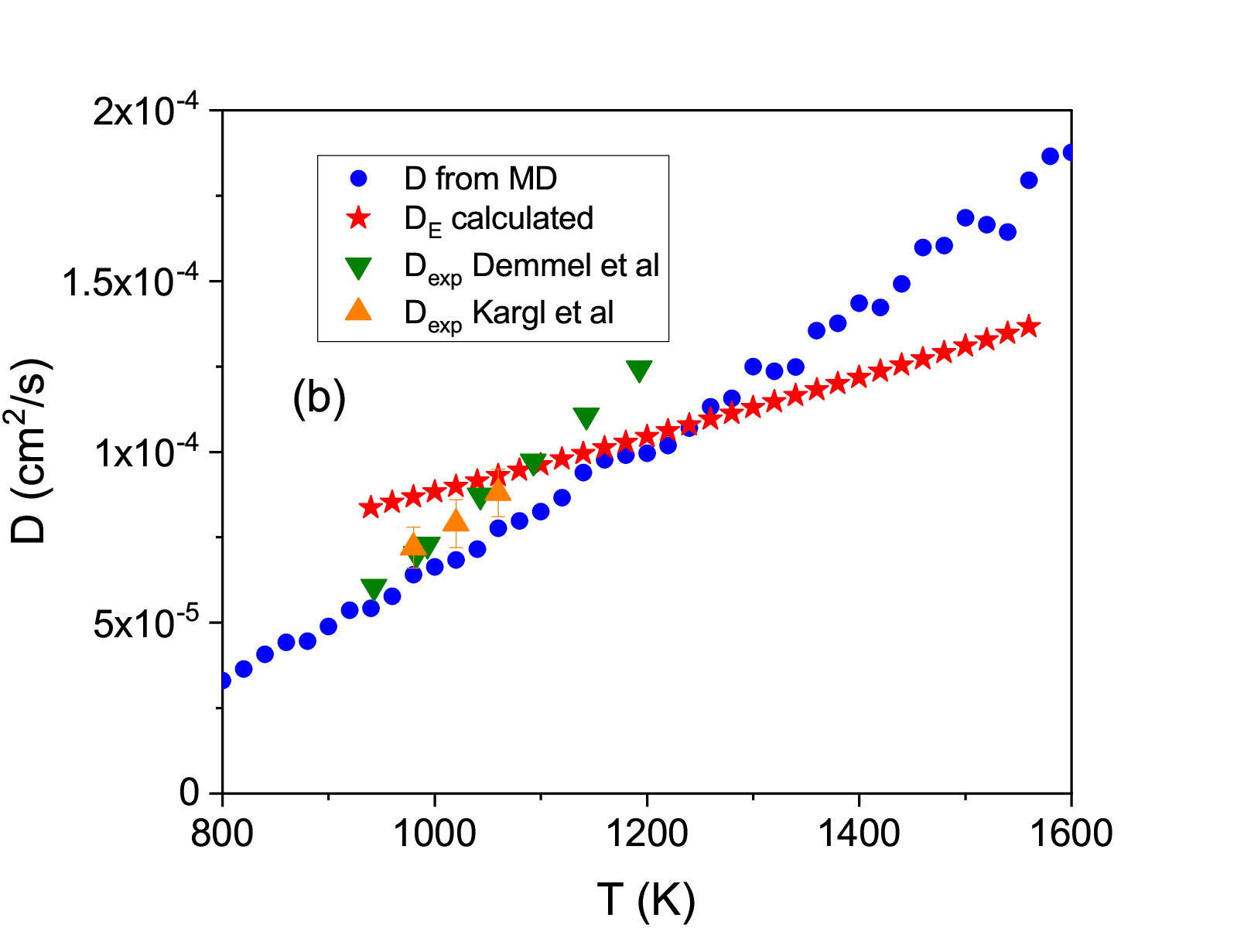}
\caption{\label{fig1} In panel (a) the Enskog diffusion coefficient $D_E$ is presented obtained from neutron scattering data (stars) together with $D$-values from the MD-simulation for Rb (circles).
Included is a calculated $D_B$ value for liquid Rb near the melting point based on only binary collisions and
a calculated $D_{theory}$ value taking into account memory effects from density fluctuations \cite{balucani1992}. In panel (b) the simulated $D$ values of Al are shown together with experimental values \cite{demmel2011,kargl2012} and calculated $D_E$ values.  }
\end{figure}
In Fig. \ref{fig1}(a) the temperature evolution of self-diffusion coefficients in liquid Rb from MD are compared to $D_E$ values derived experimentally from the linewidth $\Gamma(Q_0)$ at the structure factor at maximum $Q_0$.
Our simulated diffusion coefficients show a good agreement with previous simulation results using classical potentials \cite{balucani1992,kahl1994,wax2001,demmel2018} and a perfect agreement with a result from an ab initio simulation near the melting point \cite{shimojo1995}.
Two additional diffusion coefficients are included, namely $D_B$, calculated by taking into account only binary collisions and neglecting all collective contributions to the diffusion process, and $D_{\rm theory}$
calculated within MCT, taking into account the additional contributions from density fluctuations \cite{balucani1992}.
The $D_B$ value \cite{balucani1992} lies very near to the derived Enskog diffusion coefficient from neutron data, while $D_{\rm theory}$ agrees perfectly with the MD data, supporting the reliability of our simulation results.

A similar comparison for Al is shown in Fig. \ref{fig1}(b), noting that this time $D_E$ values are calculated from theory.
The simulated diffusion coefficients perfectly agree with previous simulated results from ab intio simulations, see the supplemental material \cite{SM}.
Also included are experimentally derived diffusion coefficients \cite{demmel2011,kargl2012}, which agree well near the melting point.
Our simulation results are in reasonable good agreement with previous first principles based simulation near the melting point
\cite{alfe1998,gonzales2002,alemany2004}. A classical simulation using an embedded atom method to derive a potential obtains
diffusion coefficients which are depending on the exact potential parameters in reasonable agreement over a wide temperature range with the results presented here \cite{kramer2010}.
With rising temperature the experimental D values lie above the simulated ones.
It remains an open question whether this difference is caused by details of the simulation potential and its evolution with density or by the analysis method of the experiment \cite{demmel2011}. A previous simulation using an embedded atom method \cite{kramer2010} suggested a slightly smaller activation energy compared to the experimental one in line with the simulations presented here, see for more details \cite{demmel2011}.

Strikingly, both Rb and Al show at low temperature $D$ values, which are smaller than the respective $D_E$ ones. However, this difference reduces with increasing temperature and a crossover is seen around $1.4~T_m$ with $D$ values becoming larger.	
\begin{figure}
\includegraphics[angle=0,width=0.48\textwidth]{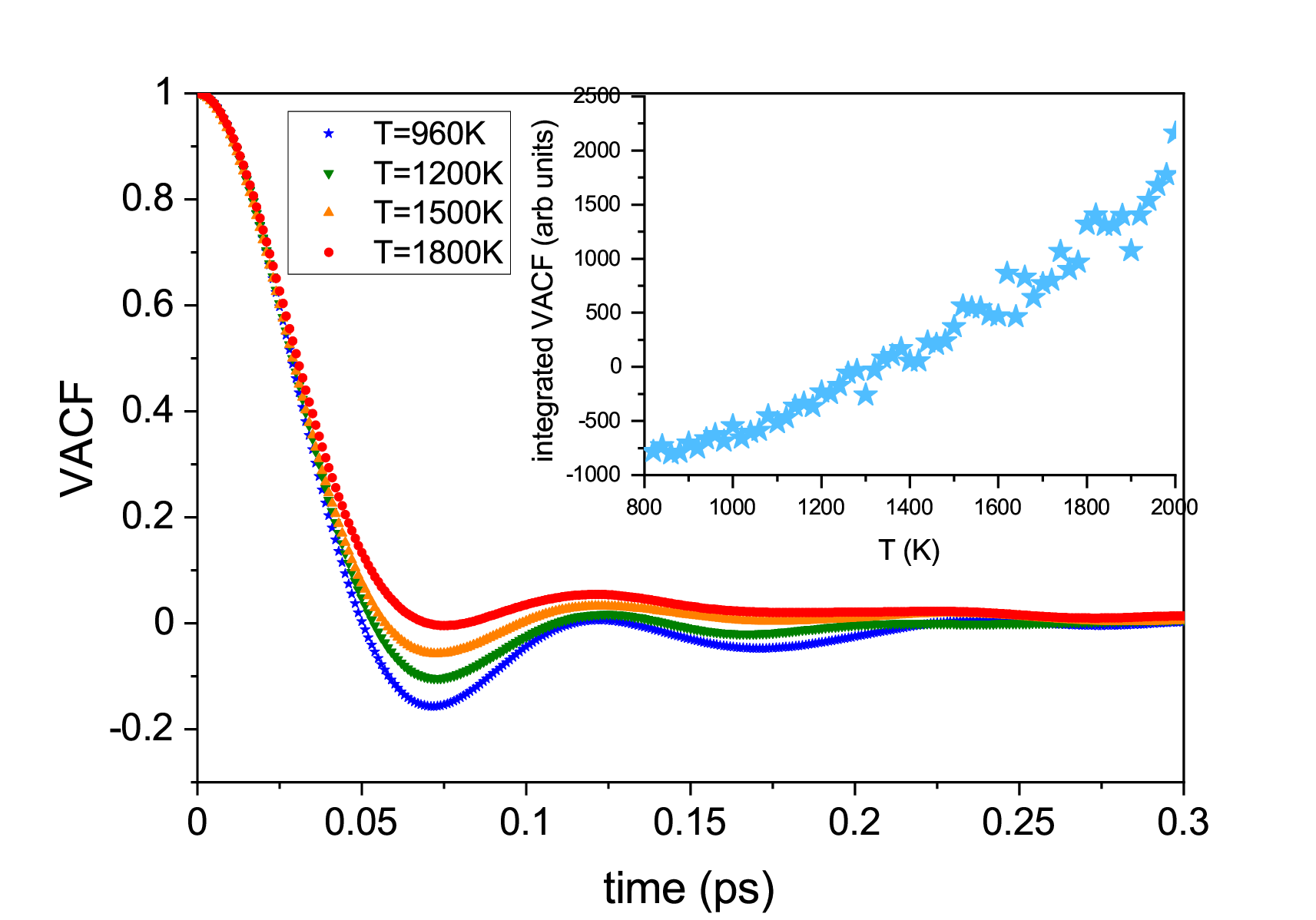}
\caption{\label{vacf1} The normalized VACF is presented for liquid Al with rising temperatures.
The inset shows the time integrated VACF between 0.04~ps and 0.24~ps for Al. }
\end{figure}	
To elucidate this behavior from a microscopic point of view, VACFs are considered.
In Fig. \ref{vacf1}, the normalized VACFs of liquid Al are displayed for several temperatures. The cage/backscattering effect is mainly expressed by the negative well around $0.07$~ps.
This leads to a reduction in the diffusion coefficient, explaining the difference between $D_E$ and $D$ near the melting point.
The negative part of the VACF completely vanishes at $T \approx 1800~K$. Changes on a more quantitative basis are shown in the inset of Fig. \ref{vacf1} through the integral $\int_{0.04}^{0.24} \psi(t) dt$ calculated over a time range that covers the most obvious changes in the velocity correlations with temperature.
Around $T \approx 1350$~K the integral crosses zero for liquid Al and becomes positive, evidencing the decreasing influence of the backscattering effect on the single particle dynamics.
The vanishing backscattering effect marks an enhanced particle mobility and leads to
a more pronounced increase of $D$ above 1400~K. This is precisely the temperature range where $D$ crosses $D_E$ as shown in Fig. \ref{fig1}.	 The corresponding results for liquid Rb are shown in the SM \cite{SM}.
\begin{figure}
\includegraphics[angle=0,width=0.48\textwidth]{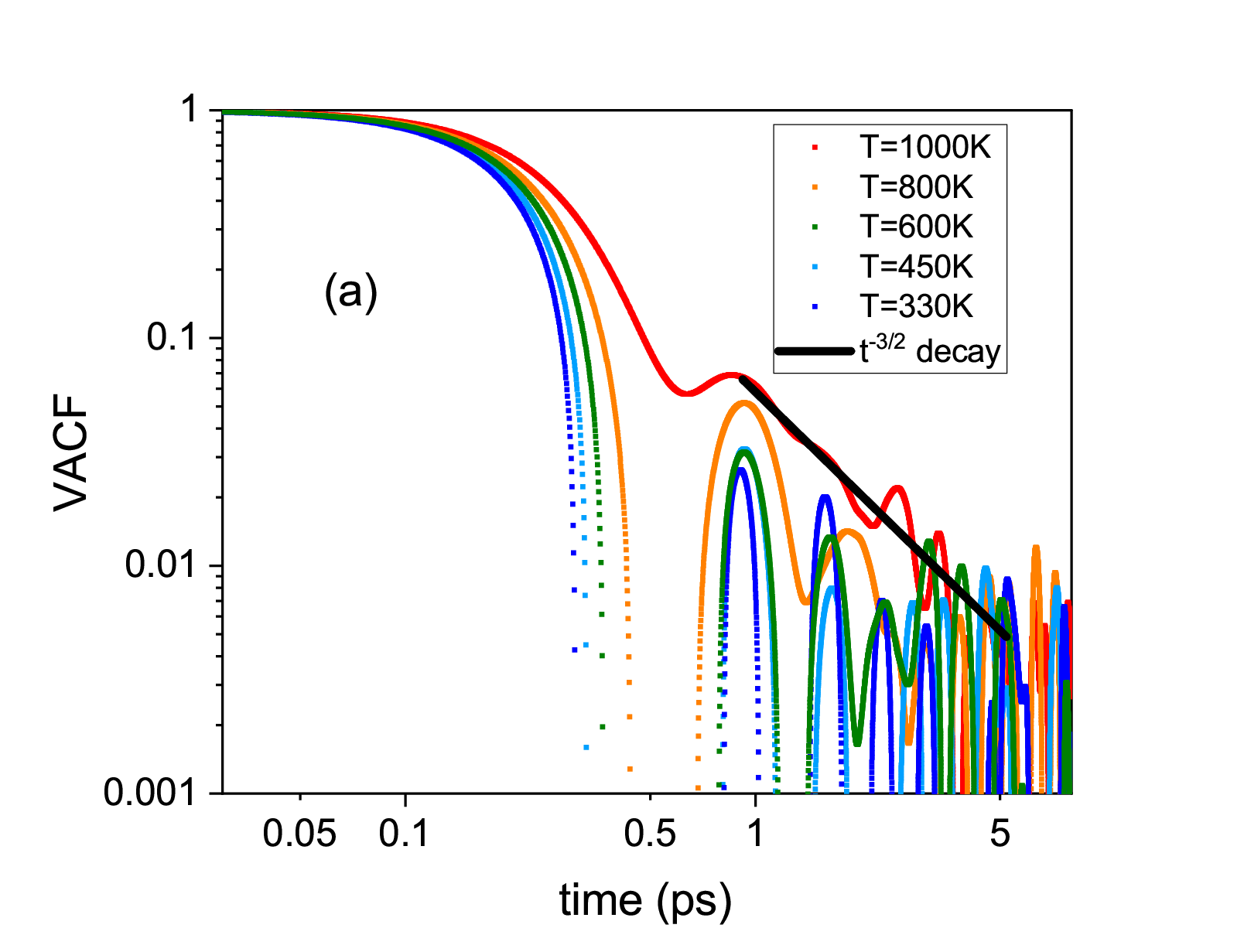}
\includegraphics[angle=0,width=0.48\textwidth]{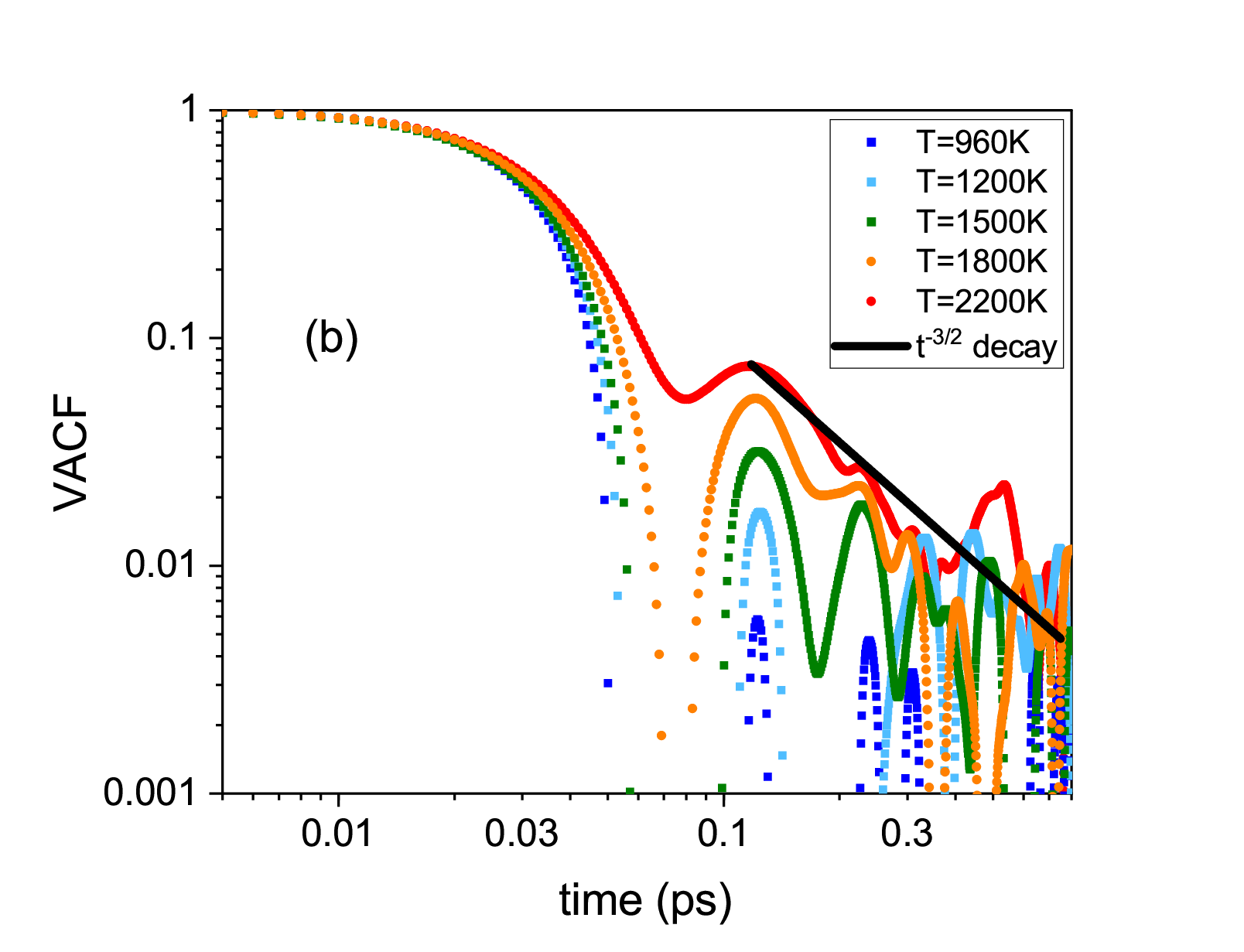}
\caption{\label{vacf2} The normalized VACF is presented for liquid Rb (a) and Al (b) on a double logarithmic scale.
The line indicates a power-law decay $VACF \propto t^{-3/2}$ for the hot fluid within a specific time range. }
\end{figure}
	
However, it is known that the coupling to density fluctuations is not the only process influencing the single particle dynamics in liquids. In the hot liquid the coupling to transverse currents plays an important role with the consequence that a power-law decay appears in the VACF.
Fig. \ref{vacf2} shows the normalized VACF for several temperatures on a
double logarithmic scale for liquid Rb and Al.
With increasing temperature the negative backscattering contribution decreases and at the highest temperature plotted the negative part has practically vanished. From previous MD simulations, it was concluded that the vortex type particle movements need about $10$ to $20$ collisions to build up \cite{alder1970,bellissima2015}.
A microscopic collision time is the Enskog time $t_E=\sqrt{m/ \pi k_B T}/ n \sigma^2 g(\sigma) $, with $n$ the particle density, $\sigma$ the hard sphere parameter and $g(\sigma)$ the pair correlation function at contact.
For Al, we obtain $t_E=0.015~ps$ and for Rb $t_E=0.08~ps$.
Hence, for times larger than about 0.15~ps for Al or larger than 0.8~ps in Rb, evidence for the LTT might be expected.
A fit with a power law $\psi(t) \propto t^{-3/2}$ is included for both liquids for their shown highest temperature (line). The fit range covers 0.15~ps to 0.5~ps for liquid Al and 0.8~ps to 5~ps for liquid Rb. In both cases, the line describes the decay of the VACF very precisely for the hot liquid. Towards lower temperature and increasing density the contribution from
the power-law decay weakens as predicted by theory and the influence of the density fluctuations overshadow
the LTT \cite{bosse1979}. Please note that the LLT contributes less than 10~\% to the correlation function.
\begin{figure}
\includegraphics[angle=0,width=0.48\textwidth]{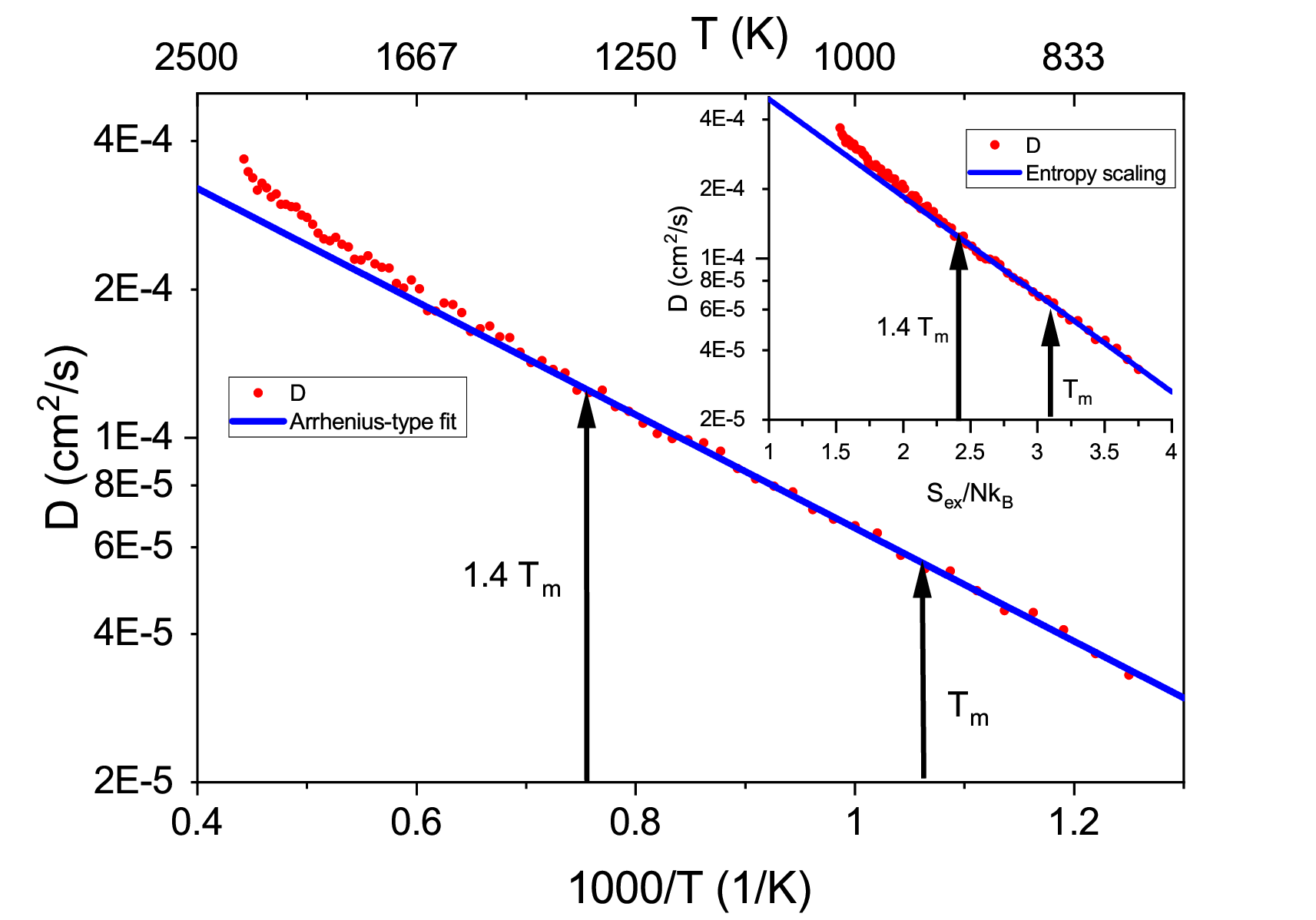}
\caption{\label{fig4} The diffusion coefficients of Al are plotted on a logarithmic scale against the inverse temperature.
Included is an Arrhenius-type fit as a line. The inset shows $D$ for Al on a logarithmic scale as a function of the reduced negative excess entropy. The line corresponds to a fit with $D ~ \exp(-S_{ex}/Nk_B)$ in the range corresponding to temperatures between $T_m$ and $1.4~T_m$.}
\end{figure}

In Fig. \ref{fig4} the diffusion coefficients of Al are plotted on a logarithmic scale against the inverse temperature.
An Arrhenius-type process describes quite well the experimental data for liquid aluminium within the limited temperature range
available \cite{demmel2011,kargl2012}.
In Fig. \ref{fig4} an Arrhenius process was fitted between $T_m$ and $1.4~T_m$, which describes the diffusion
coefficients very precisely within this temperature range.
However, at higher temperatures, the diffusion coefficients deviate to a faster particle dynamics compared to the dynamics expected from the Arrhenius-type behavior. The deviation to a non-Arrhenius behavior starts around
the same temperature range, where $D$ crosses the $D_E$ value, see Fig. \ref{fig1}.
A similar behavior is observed for Rb, see the SM \cite{SM}.
	
The crossover when $D$ becomes equal to $D_E$ at a temperature range around $1.4~T_m$ mirrors the collective dynamics behavior studied previously over a wide temperature range in liquid Rb and Al \cite{demmel2006a,demmel2015,demmel2021}.
A crossover was observed in the structural relaxation time and the generalized longitudinal viscosity in the same temperature range, providing strong support to a scenario when diffusion gradually phase out from a regime governed by density fluctuations to one overtaken by transverse currents.
Further evidence for a distinct change in dynamics within the equilibrium liquid state was obtained for liquid Ga and Pb \cite{demmel2008,demmel2020}. The analysis of experimental data suggested that with decreasing temperature an additional slow relaxation process appears in the dynamics of liquid rubidium at the structure factor maximum \cite{demmel2012,demmel2018}.
Previously, that relaxation process was interpreted as an essential process towards structural arrest \cite{balucani1989}, and it turned out that this process is the important contribution to the mode coupling description of the single particle motion for describing the cage effect near the melting point \cite{balucani1990,balucani1992}. Upon cooling the appearance of this slow structural relaxation process coincides with the slowing down of the diffusion coefficient due to the cage effect.
In the same temperature range the slope of the shear viscosity changes strongly. The decreasing viscosity with rising temperature enables an increase of the lifetime of shear relaxation modes, what might support the building up of vortex movements.
	
In the supercooled region a further non-Arrhenius behavior is observed, which is related to structural arrest towards the glass transition, For example, a non-Arrhenius departure was observed for diffusion of Ni in glass forming alloys \cite{meyer2002}.
For monatomic liquid metals the non-Arrhenius behavior occurs below the melting point in the supercooled liquid
and at the lowest temperature evidence for such a deviation can be seen in Fig. 3 of the SM. A similar observation was made recently for liquid Zr \cite{becker2020}.
A non-Arrhenius behavior can be interpreted that the activation energy for a diffusive step is changing and can be put into the context of an energy landscape picture \cite{debene2001}. For liquid Al the inherent structural energy of an atom in the potential energy landscape was studied over a wide temperature range \cite{demmel2021}.
A crossover in the inherent structural energy was observed around $T=1300~K$ when the liquid enters a free fluid-like state, whereas below this temperature range the particle sinks deeper into an energy basin with an eventual arrest at the glass state.
Our observations agree with this picture.
	
A different viewpoint how the diffusion process changes with rising temperature was put forward with a direct connection between particle mobility and accessible configurations of a system enshrined in the excess entropy $S_{ex}$, the difference between the total entropy of the liquid and the entropy of an ideal gas \cite{rosenfeld1977,dzugutov1996}.
This proposal led to a scaling law  $D ~ \exp(-B S_{ex}/N k_B)$ where $B$ should be a universal parameter.
The inset in figure \ref{fig4} shows the diffusion coefficients against the reduced negative excess entropy
on a logarithmic scale for liquid Al.
The fitted line for $S_{ex}$ indicates a good agreement for a wide range of entropy values.
Within MD-simulations, such an entropy scaling law was confirmed for many liquids over a wide range in temperatures \cite{dzugutov1996,jakse2021,jakse2013b,jakse2016a}. Nevertheless, for small excess entropy or respective high temperatures, a deviation from the scaling law is obvious. This enhancement of the diffusion coefficients from the scaling law was previously noted and related to a crossover from cage diffusion to vortex diffusion \cite{dzugutov1996}. The here presented results confirm this assumption. The corresponding results for liquid Rb are plotted in the supplemental material \cite{SM}.
	
In summary, a holistic view on the microscopic self-dynamics of liquid metals with rising temperature was obtained from extensive MD simulations. Diffusion coefficients of liquid Al and Rb are crossing the Enskog $D_E$ values in a temperature range around $1.4~T_m$, indicating a profound change in the diffusion mechanism.
The associated VACFs demonstrate a decrease of the cage effect with a gradual set-in of the LTT, signaling a competition of density fluctuations with vortex particle patterns from transverse currents.
That change in underlying collective dynamics is at the origin of the non-Arrhenius behavior of the diffusion coefficient $D$ at temperatures above $1.4~T_m$. A crossover in collective particle dynamics was observed previously within this temperature range and the single particle dynamics mirrors these underlying changes in dynamics.
Our detailed investigation relates the changes in the potential energy landscape with the diffusion dynamics and provides an explanation for the deviations in the scaling law of the configurational entropy in the hot fluid.
The here presented MD results could serve as a basis for a complete single particle diffusion theory, which encompasses the whole density range of a fluid.
	
We acknowledge the CINES and IDRIS under Project No. INP2227/72914, as well as
CIMENT/GRICAD for computational resources. This work has been partially supported by MIAI@Grenoble Alpes (ANR-19-P3IA-0003) and international SOLIMAT projet ANR-22-CE92-0079-01. Discussions within the French collaborative network in artificial intelligence in materials science GDR CNRS 2123 (IAMAT) is also acknowledged. Support by the Science and Technology Facilities Council is acknowledged. FD is grateful to Chr. Morkel for numerous discussions on this subject.

\end{document}